\documentclass[twocolumn,showpacs,preprintnumbers,amsmath,amssymb,superscriptaddress,prb]{revtex4}

\usepackage{graphicx}
\usepackage{dcolumn}
\usepackage{hyperref}
\usepackage{color}
\usepackage{bm}

\usepackage[normalem]{ulem}

\usepackage{t1enc}
\usepackage[polish,english]{babel}
\usepackage[cp1250]{inputenc}
\usepackage{polski}
\usepackage[T1]{fontenc}
\begin{document}

\title{Spin effects in transport through single-molecule magnets in the sequential and cotunneling regimes}

\author{Maciej Misiorny}
 \email{misiorny@amu.edu.pl}
\affiliation{Faculty of Physics, Adam Mickiewicz University,
61-614 Pozna\'{n}, Poland}

\author{Ireneusz Weymann}
 \email{weymann@amu.edu.pl}
\affiliation{Faculty of Physics, Adam Mickiewicz University,
61-614 Pozna\'{n}, Poland} \affiliation{Department of Theoretical
Physics, Institute of Physics, Budapest University of Technology
and Economics, H-1521 Budapest, Hungary}

\author{J\'{o}zef Barna\'{s}}
\email{barnas@amu.edu.pl} \affiliation{Faculty of Physics, Adam
Mickiewicz University, 61-614 Pozna\'{n}, Poland}
\affiliation{Institute of Molecular Physics, Polish Academy of
Sciences, 60-179 Pozna\'{n}, Poland
}%

\date{\today}

\begin{abstract}
We analyze the stationary spin-dependent transport through a
single-molecule magnet weakly coupled to external ferromagnetic
leads. Using the real-time diagrammatic technique, we calculate
the sequential and cotunneling contributions to current, tunnel
magnetoresistance and Fano factor in both linear and nonlinear
response regimes.  We show that the effects of cotunneling are
predominantly visible in the blockade regime and lead to
enhancement of tunnel magnetoresistance (TMR) above the Julliere
value, which is accompanied with super-Poissonian shot noise due
to bunching of inelastic cotunneling processes through different
virtual spin states of the molecule. The effects of external
magnetic field and the role of type and strength of exchange
interaction between the LUMO level and the molecule's spin are
also considered. When the exchange coupling is ferromagnetic, we
find an enhanced TMR, while in the case of antiferromagnetic
coupling we predict a large negative TMR effect.
\end{abstract}



\maketitle

\section{Introduction}

Owing to recent advances in experimental techniques, it is now
possible to study transport properties of individual nanoscale
objects, like quantum dots,~\cite{vanderWiel_RevModPhys75/02}
nanotubes,\cite{Tans_Nature386/97,Tans_Nature393/98,
Tsukagoshi_Nature401/99,Cleuziou_NatNanotech1/06} and other
molecules.
\cite{Reed_Science278/97,Porath_PRB56/97,Porath_Nature403/00,Park_Nature407/00,
Reichert_PRL88/02,Grose_condmat08} Investigation of electron
transport through molecules is stimulated by the prospect of a new
generation of molecule-based electronic and spintronic devices. It
turns out that owing to their unique optical, magnetic and
mechanical properties, molecules are ideal candidates for
constructing novel hybrid devices of functionality which  would be
rather hardly accessible in the case of conventional silicon-based
electronic systems.
\cite{Joachim_Nature408/00,Nitzan_Science300/03,
Tao_NatNanotech1/06,Bogani_NatureMater7/08} For instance, one
interesting feature of nanomolecular systems, which does not have
counterpart in bulk materials, concerns the interplay between the
quantized electronic and mechanical degrees of
freedom.~\cite{Park_Nature407/00}

In this paper we deal with one specific class of molecules which
possess an intrinsic magnetic moment, referred to as
single-molecule magnets (SMMs).
\cite{Caneschi_JMMM200/99,Christou_MRSBull25/00,Gatteschi_book}
Such molecules are characterized by a significant Ising-like
magnetic anisotropy and a high spin number $S$, which give rise to
an energy barrier that the molecule has to overcome to reverse its
spin orientation. At higher temperatures, the SMM's spin can
freely rotate, whereas below a certain temperature it becomes
trapped in one of two metastable orientations. Since magnetic
bistability is one of the key properties to be utilized in
information processing technologies, SMMs have attracted much
attention and a great deal of effort was undertaken to measure
electronic transport through a
SMM.~\cite{Heersche_PRL96/06,Ni_APL89/06,Jo_NanoLett6/06,Henderson_JApplPhys101/07,Voss_PRB78/08}
The experiments carried out to date have concerned only the case
of SMMs coupled to nonmagnetic electrodes. However, it has been
suggested recently that spin-polarized currents (when the leads
are ferromagnetic, for instance) can be used to manipulate the
magnetic state of a SMM.
\cite{Misiorny_PRB75/07,Misiorny_PRB76/07,
Timm_PRB73/06,Elste_PRB73/06,Misiorny_PSS_FA} Such a
current-induced magnetic switching (CIMS) of a SMM takes place as
a consequence of the angular momentum transfer between the
molecule and conduction electrons.

When considering coupling strength between the molecule and
external leads, one can generally distinguish between weak and
strong coupling regimes. In the latter case, i.e. when resistance
of the contact between the molecule and electrodes becomes smaller
than the quantum resistance, the electronic correlations may lead
to formation of the Kondo effect.
\cite{Romeike_PRL96I/06,Romeike_PRL97/06,
Leuenberger_PRL97/06,Roosen_PRL100/08,Gonzalez_PRB78/08} These
correlations result in a screening of the SMM's spin by conduction
electrons of the leads, giving rise to a peak in the density of
states and full transparency through the molecule. On the other
hand, in the weak coupling regime, the Coulomb correlations lead
to blockade phenomena. \cite{Grabert_book_1992} For voltages lower
than a certain threshold value, sequential tunneling processes
through the molecule are then exponentially suppressed due to
Coulomb correlations and/or size quantization. However, once the
bias voltage exceeds the threshold value, the electrons can tunnel
one-by-one through the molecule. The latter regime is known as the
sequential tunneling regime, and the former one is often referred
to as the Coulomb blockade or cotunneling regime.
\cite{Averin_PLA140/89,Averin_PRL65/90} It should be noted,
however, that although the sequential processes are suppressed in
the Coulomb blockade regime, current still can flow due to second-
and higher-order tunneling processes, which involve correlated
tunneling through virtual states of the molecule. Furthermore,
although higher-order processes play a substantial role mainly in
the cotunneling regime, they remain active in the whole range of
transport voltages, especially on resonance, leading to
renormalization of the molecule levels and smearing of the Coulomb
steps. \cite{Koenig_PhD} Therefore a suitable theoretical method
should be used to properly investigate transport through molecules
in the regime where both the sequential and cotunneling processes
coexist and determine transport properties. The existing
analytical studies of electronic transport through SMMs in the
weak coupling regime were based on the standard perturbation
approach, \cite{Misiorny_PRB76/07,Timm_PRB73/06,Misiorny_PSS_FA,
Elste_PRB73/06,Misiorny_IEEETransMag,Misiorny_cotunneling} and
they dealt separately either with the sequential or cotunneling
regime, with one attempt of combining them.~\cite{Elste_PRB75/07}
Nevertheless, to properly take into account the nonequilibrium
many-body effects such as for example on-resonance level
renormalization or level splitting due to an effective exchange
field, simple rate equation arguments are not sufficient.

The main objective of the present paper is thus a systematic
analysis of charge and spin transport through a SMM. This has been
achieved by employing the real-time diagrammatic technique,
\cite{Schoeller_PRB50/94} which enables accurate study of
transport properties in the {\it full} weak coupling regime. In
particular, including the first- and second-order self-energy
diagrams, we calculate the current, tunnel magnetoresistance (TMR)
and shot noise in the presence of sequential tunneling,
cotunneling and cotunneling-assisted sequential tunneling
processes. We show that the second-order processes determine
transport in the Coulomb blockade regime, leading for instance to
enhanced tunnel magnetoresistance effect as compared to the value
based on the Julliere model, \cite{Julliere_PLA54/75} and to
super-Poissonian shot noise due to bunching of inelastic
cotunneling processes through the molecule. In addition, we also
discuss the effects due to external magnetic field as well as the
role of strength and type of  exchange interaction between the
molecule's spin and conduction electrons.

The paper is organized as follows. In section II we describe the
model of a single-molecule magnet coupled to ferromagnetic leads.
The real-time diagrammatic technique used in calculations is
presented briefly in section III. Section IV is devoted to
numerical results and their discussion. In particular, the
conductance, tunnel magnetoresistance and shot noise in the linear
and nonlinear response regimes are analyzed in subsections A and
B, respectively. The dependence of transport properties on the
strength of exchange coupling is discussed in subsection C, while
the effects of longitudinal external magnetic field are considered
in subsection D. Furthermore, we also briefly discuss transport
characteristics in the case when the exchange coupling is
antiferromagnetic, subsection E. Finally, conclusions are given in
section V.

\section{\label{Sec:Model}Description of model}

\begin{figure}
    \includegraphics[width=0.85\columnwidth]{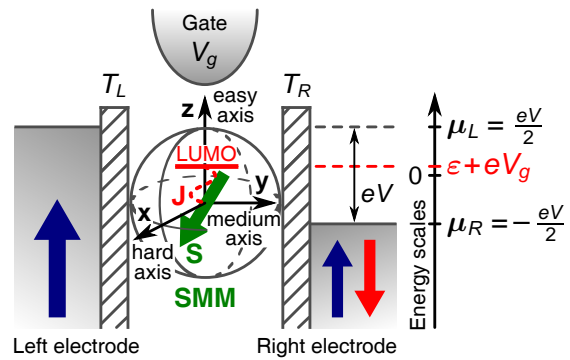}
    \caption{\label{Fig1} (color online)
    Schematic representation of the system under consideration.
    The system consists of a SMM weakly coupled to two ferromagnetic
    electrodes with the collinear configuration of their magnetic moments,
    i.e. either parallel or antiparallel.  Due to symmetrically
    applied bias voltage $V=(\mu_L-\mu_R)/e$, where $\mu_{L(R)}$
    denotes the electrochemical potential of the left (right) lead,
    the LUMO level is independent of $V$.
    Position of the LUMO level, however, can be tuned by the gate voltage $V_g$.}
\end{figure}

In this paper we consider a model SMM which is attached to two
metallic ferromagnetic electrodes, see Fig.~\ref{Fig1}. The
molecule is assumed to be weakly coupled to the leads, whose
magnetizations form a collinear configuration, either parallel or
antiparallel. The limit of strong coupling, where interesting
phenomena such as the Kondo effect \cite{Romeike_PRL96I/06,Romeike_PRL97/06,
Leuenberger_PRL97/06,Roosen_PRL100/08,Gonzalez_PRB78/08} can be
observed, is not considered here.

Electronic transport through the molecule is assumed to take place
only \emph{via} the lowest unoccupied molecular orbital (LUMO) of
the SMM, which is coupled to the internal magnetic core of the
molecule \emph{via}  exchange interaction.  Moreover, we also
neglect all other unoccupied levels which are assumed to be well
above the LUMO level and therefore cannot take part in transport
for voltages of interest.~\cite{Misiorny_PSS_FA} Furthermore,
following previous theoretical
studies,~\cite{Misiorny_PRB75/07,Misiorny_PRB76/07,
Timm_PRB73/06,Elste_PRB73/06} we restrict our considerations to
the case of molecules with vanishingly small transverse
anisotropy.

Taking the above into account, a SMM coupled to external leads can
be described by Hamiltonian of the general form
\begin{equation}\label{eq:Hamiltonian}
    \mathcal{H}=\mathcal{H}_\textrm{SMM}+\mathcal{H}_\textrm{leads}+\mathcal{H}_\textrm{T}
    \,.
\end{equation}
The first term on the right hand side describes the SMM and is
assumed in the following form:
\begin{align}\label{eq:Hamiltonian_SMM}
    \mathcal{H}_\textrm{SMM} =& -\Big[D +\sum_{\sigma}D_{1}\, c_\sigma^\dag
    c_\sigma^{} + D_{2}\, c_\uparrow^\dag c_\uparrow^{} c_\downarrow^\dag c_\downarrow^{}
    \Big]S_z^2
    \nonumber\\
    &+ \sum_{\sigma} \varepsilon\,c_\sigma^\dag c_\sigma^{}
    + U\, c_\uparrow^\dag c_\uparrow^{} c_\downarrow^\dag c_\downarrow^{}
    -J \textbf{s}\cdot\textbf{S}
    \nonumber\\
    &+g\mu_\textrm{B}(S_z+s_z)H_z.
\end{align}
The first line of Eq.~(\ref{eq:Hamiltonian_SMM}) accounts for the
uniaxial magnetic anisotropy of a SMM, characterized by the
uniaxial anisotropy constant $D$ of a free-standing (neutral)
molecule. When a bias voltage is applied, the LUMO level can be
charged with up to two electrons, which in turn can affect the
magnitude of the uniaxial anisotropy. The relevant corrections are
included by the constants $D_1$ and $D_2$. Moreover, $S_z$ denotes
the $z$ component of the internal (core) spin operator
$\textbf{S}$, whereas $c_\sigma^\dag$ $(c_\sigma^{})$ is the
creation (annihilation) operator of an electron in the LUMO level.
We note that the Hamiltonian (2) is applicable to situations where
electronic structure of the molecules's magnetic core is not
changed by adding one or two electrons to the LUMO level, except
for modification of the anisotropy constants.

The second line of the Hamiltonian $\mathcal{H}_\textrm{SMM}$
describes the LUMO level of energy $\varepsilon$, with $U$ being
the Coulomb energy of two electrons of opposite spins that can
occupy this level. Although the position of the LUMO level can be
modified by the gate voltage $V_g$, it remains independent of the
symmetrically applied bias voltage $V$. An important term for the
present discussion is the last one, given explicitly by
\begin{equation}
    J\textbf{s}\cdot\textbf{S}=\frac{J}{2}c_\uparrow^\dag c_\downarrow^{}S_-
    + \frac{J}{2} c_\downarrow^\dag c_\uparrow^{}
    S_+ + \frac{J}{2} \Big[c_\uparrow^\dag
    c_\uparrow^{}-c_\downarrow^\dag c_\downarrow^{}\Big]S_z \,,
\end{equation}
which stands for exchange coupling between the magnetic core of a
SMM, represented by the spin $\textbf{S}$, and electrons in the
LUMO level, described by the local spin operator
$\textbf{s}=\frac{1}{2}\sum_{\sigma\sigma'}c_\sigma^\dag
\bm{\sigma}_{\sigma\sigma'} c_{\sigma'}$, where $\bm{\sigma}$ is
the vector of Pauli matrices. This interaction can be either of
ferromagnetic ($J>0$) or antiferromagnetic ($J<0$) type. Finally,
the last term of $\mathcal{H}_\textrm{SMM}$ describes the Zeeman
splitting associated with the magnetic field applied along the
easy axis of the molecule, where $g$ stands for the Land\'{e}
factor, and $\mu_\textrm{B}$ is the Bohr magneton.

In general, the molecular Hamiltonian $\mathcal{H}_\textrm{SMM}$
is not diagonal, except for the case of a free-standing
(uncharged) uniaxial SMM. It has been
shown~\cite{Timm_PRB73/06,Misiorny_PRB76/07,Misiorny_PSS_FA} that
for the molecules with no transverse anisotropy,
$\mathcal{H}_\textrm{SMM}$ commutes with the $z$th component
($S_t^z$) of the total spin $\textbf{S}_t\equiv
\textbf{S}+\textbf{s}$, hence allowing us to analytically
diagonalize it in the basis represented by the eigenvalues $m$ of
$S_t^z$ and the corresponding occupation number $n$ of the LUMO
level. In a general case, on the other hand, the problem can be
dealt with numerically by performing a unitary transformation
$U^\dag\mathcal{H}_\textrm{SMM}U =
\widetilde{\mathcal{H}}_\textrm{SMM}$ to a new basis in which
$\widetilde{\mathcal{H}}_\textrm{SMM}$ is diagonal. Consequently,
we obtain the set of relevant eigenvectors $|\chi\rangle$ and the
corresponding eigenvalues $\varepsilon_\chi$ satisfying
$\widetilde{\mathcal{H}}_\textrm{SMM}| \chi\rangle =
\varepsilon_\chi |\chi\rangle$.

The second term of Eq.~(\ref{eq:Hamiltonian}) describes
ferromagnetic electrodes, and the $q$th electrode ($q=L,R$) is
characterized by noninteracting itinerant electrons with the
dispersion relation $\varepsilon_{\textbf{k}\sigma}^q$, where
$\textbf{k}$ denotes a wave vector and $\sigma$ is the electron's
spin. As a result, the lead Hamiltonian can be written as
\begin{equation}
    \mathcal{H}_\textrm{leads}=\sum_q\sum_{\textbf{k}, \sigma}
    \varepsilon_{\textbf{k}\sigma}^q\: a_{\textbf{k}\sigma}^{q\dag}
    a_{\textbf{k}\sigma}^q \,,
\end{equation}
where $a_{\textbf{k}\sigma}^{q\dag}$ ($a_{\textbf{k}\sigma}^q$) is
the creation (annihilation) operator for an electron in the $q$th
electrode. The degree of spin polarization of the ferromagnetic
lead $q$ can be described by the parameter $P_q$,
$P_q=(D_+^q-D_-^q)/(D_+^q+D_-^q)$, with $D_\pm^q$ denoting the
density of states for majority (upper sign) and minority (lower
sign) electrons at the Fermi level in the lead $q$.

Finally, the last term $\mathcal{H}_\textrm{T}$ of the total
Hamiltonian~(\ref{eq:Hamiltonian}) describes tunneling processes
between the molecule and the leads, and it is given by
\begin{equation}
    \mathcal{H}_\textrm{T}=\sum_q\sum_{\textbf{k},\sigma}
    \Big[T_q\, a_{\textbf{k}\sigma}^{q\dag}c_\sigma^{} +
    T_{q}^* c_\sigma^\dag a_{\textbf{k}\sigma}^q\Big] \,,
\end{equation}
with $T_q$ denoting the tunnel matrix element between the molecule
and the $q$th lead. Due to the tunneling processes, the LUMO level
of the molecule acquires a finite spin-dependent width,
$\Gamma_\sigma =\sum_q \Gamma_\sigma^q$, where
$\Gamma_\sigma^q=2\pi|T_q|^2D_\sigma^q$. The parameters
$\Gamma^{q}_{\pm}$ can be also expressed in terms of the spin
polarization $P_q$ of the lead $q$ as
$\Gamma^{q}_{\pm}=\Gamma_{q}(1\pm P_q)$ for spin-majority (upper
sign) and spin-minority (lower sign) electrons, where
$\Gamma_{q}=(\Gamma^{q}_{+}+\Gamma^{q}_{-})/2$. In the following
these parameters will be used to describe the strength of coupling
between the LUMO level and the leads. Unless stated otherwise, the
couplings are assumed to be symmetric, $\Gamma_{\rm L}=\Gamma_{\rm
R}=\Gamma/2$.

\section{Method of calculations}

Among different available methods, only a few enable us to analyze
spin-dependent transport of the considered system in both the
sequential and Coulomb blockade regimes within one fully
consistent theoretical approach.~\cite{Timm_PRB77/08} In
particular, here, we employ the real-time diagrammatic
technique,~\cite{Schoeller_PRB50/94,Koenig_PRB54/96,
Koenig_PhD,Thielmann_PRB68/03,Thielmann_PRL95/05,
Weymann_PRB72/05_TMR} which has already proven its reliability and
versatility in studying transport properties of various nanoscopic
systems.

The basic idea of this technique relies on a systematic
perturbation expansion of the reduced density matrix of the system
under discussion and the operators of interest with respect to the
coupling strength $\Gamma$ between the LUMO level and the leads.
All quantities, such as the current $I$, differential conductance
$G$ and the (zero-frequency) current noise $S$ are essentially
determined by the nonequilibrium time evolution of the reduced
density matrix for the molecule's degrees of freedom. In the case
considered in this paper, the density matrix has only diagonal
matrix elements, $p_{\chi}(t)$, which correspond to probability of
finding the molecule in state $|\chi\rangle$ at time $t$.
Following the matrix notation introduced by Thielmann \emph{et
al.},~\cite{Thielmann_PRB68/03} the vector $\textbf{p}(t)$ of the
probabilities is given by the
relation~\cite{Schoeller_PRB50/94,Koenig_PRB54/96,Koenig_PhD}
\begin{equation}
    \textbf{p}(t)=\bm{\Pi}(t,t_0)\textbf{p}(t_0) \,,
\end{equation}
where $\bm{\Pi}(t,t_0)$ is the propagator matrix whose elements,
$\Pi_{\chi'\chi}(t,t_0)$, describe the time evolution of the
system that propagates from a state $|\chi\rangle$ at time $t_0$
to a state $|\chi'\rangle$ at time $t$, and $\textbf{p}(t_0)$ is a
vector representing the distribution of initial probabilities. In
principle, the whole dynamics of the system is governed by the
time evolution of the reduced density matrix. Furthermore, this
time evolution can be schematically depicted as a sequence of
irreducible diagrams on the Keldysh contour, \cite{Koenig_PhD}
which after summing up correspond to irreducible self-energy
blocks $W_{\chi'\chi}(t',t)$. \cite{Thielmann_PRB68/03} The
self-energy matrix $\textbf{W}(t',t)$ is therefore one of the
central quantities of the real-time diagrammatic technique, as its
elements $W_{\chi'\chi}(t',t)$ can be interpreted as generalized
transition rates between two arbitrary molecular states:
$|\chi\rangle$ at time $t$ and $|\chi'\rangle$ at time $t'$.
Consequently, the Dyson equation for the propagator is obtained in
the form
\cite{Koenig_PhD,Thielmann_PRB68/03,Schoeller_PRB50/94,Koenig_PRB54/96}
\begin{equation}\label{eq:Dyson_equation}
    \bm{\Pi}(t,t_0)=\textbf{1}+\int_{t_0}^{t}\textrm{d}t_2\int_{t_0}^{t_2}\textrm{d}t_1
    \textbf{W}(t_2,t_1)\bm{\Pi}(t_1,t_0) \,.
\end{equation}
By multiplying Eq.~(\ref{eq:Dyson_equation}) from the right hand
side with $\textbf{p}(t_0)$, and differentiating it with respect
to time $t$, one gets the general kinetic equation for the
probability vector $\textbf{p}(t)$,
\begin{equation}
    \frac{\textrm{d}}{\textrm{d}t}\textbf{p}(t) =
    \int_{t_0}^{t}\textrm{d}t_1 \textbf{W}(t,t_1)\textbf{p}(t_1) \,.
\end{equation}
In the limit of stationary transport the aforementioned formula
reduces to the steady state master-like equation
\cite{Schoeller_PRB50/94,Thielmann_PRB68/03,Koenig_PRB54/96,Koenig_PhD}
\begin{equation} \label{eq:stationary_probabilities}
   \Big(\widetilde{\textbf{W}}\textbf{p}^\textrm{st}\Big)_\chi=\Gamma \delta_{\chi\chi_{0}^{}} \,,
\end{equation}
where $\textbf{p}^\textrm{st}=\lim_{t\rightarrow
\infty}\textbf{p}(t)=\lim_{t_0\rightarrow -\infty}\textbf{p}(0)$
is the stationary probability vector, independent of initial
distribution. On the other hand, $\widetilde{\textbf{W}}$ denotes
the Laplace transform of the self-energy matrix
$\textbf{W}(t',t)$, whose one arbitrary row $\chi_0^{}$ has been
replaced with $(\Gamma,\ldots,\Gamma)$ to include the
normalization condition for the probabilities $\sum_\chi
p_\chi^\textrm{st}=1$. Knowing the probabilities, the electric
current flowing through the system can be calculated from the
formula \cite{Thielmann_PRB68/03}
\begin{equation} \label{eq:current}
    I=\frac{e}{2\hbar} \textrm{Tr}\left[\textbf{W}^{I}\textbf{p}^{st}\right] \,,
\end{equation}
where the matrix $\textbf{W}^{I}$ denotes the self-energy matrix
in which one {\it internal} vertex originating from the expansion
of tunneling Hamiltonian $\mathcal{H}_\textrm{T}$ has been
substituted with an {\it external} vertex for the current
operator.

In order to calculate the transport quantities in both the deep
Coulomb blockade and the sequential tunneling regime in each order
in tunneling processes, we perform the perturbation expansion in
$\Gamma$ adopting the so-called {\it crossover} perturbation
scheme, \cite{Weymann_PRB72/05_TMR} i.e. we expand the self-energy
matrices,
$\widetilde{\textbf{W}}=\sum_{n=1}^{\infty}\widetilde{\textbf{W}}^{\raisebox{-4pt}
{\scalebox{0.7}{(n)}}}$ and $\textbf{W}^I =
\sum_{n=1}^{\infty}\textbf{W}^{I(n)}$. Here, the first order of
expansion ($n=1$) corresponds to sequential tunneling processes,
while the second-order contribution ($n=2$) is associated with
cotunneling processes. In the present calculations we take into
account both the first- and second-order diagrams, which allows us
to resolve the transport properties in the {\it full} weak
coupling regime, i.e. in the cotunneling as well as in the
sequential tunneling regimes. Furthermore, by considering the
$n=1$ and $n=2$ terms of the expansion, we systematically include
the effects of LUMO level renormalization, cotunneling-assisted
sequential tunneling, as well as effects associated with an
exchange field exerted by ferromagnetic leads on the molecule.
\cite{Koenig_PRL90/03,Weymann_PRB72/05_TMR,Weymann_PRB75/07} For
$n\leq 2$, the stationary probabilities can be found from
Eq.~(\ref{eq:stationary_probabilities}), with
$\widetilde{\textbf{W}} = \widetilde{\textbf{W}}^{\raisebox{-4pt}
{\scalebox{0.7}{(1)}}} + \widetilde{\textbf{W}}^{\raisebox{-4pt}
{\scalebox{0.7}{(2)}}}$. On the other hand, the current is
explicitly given by Eq.~(\ref{eq:current}) where one has to take
$\textbf{W}^{I} = \textbf{W}^{I(1)} + \textbf{W}^{I(2)}$. The key
problem is now the somewhat lengthy but straightforward
calculation of the respective self-energy matrices, which can be
done using the corresponding diagrammatic rules.
\cite{Schoeller_PRB50/94,Koenig_PRB54/96,
Koenig_PhD,Thielmann_PRB68/03,Weymann_PRB72/05_TMR} An example of
explicit formula for a second-order self-energy between arbitrary
states $|\chi\rangle$ and $|\chi'\rangle$ can be found in
Ref.~[\onlinecite{Weymann_PRB78/08}].

With recent progress in detection of ultra-small signals, it has
become clear that the information about the system transport
properties can also be extracted from the measurement of current
noise.~\cite{Blanter_PhysRep336/00} In fact, the shot noise
contains information about various correlations, coupling
strengths, effective charges, etc., which is sometimes
unaccessible just from measurements of electric current.
Therefore, to make the analysis more self-contained, in this paper
we will also calculate and discuss the zero-frequency shot noise.
The shot noise is usually defined as the correlation function of
the current operators, and its Fourier transform in the limit of
low frequencies is given by~\cite{Blanter_PhysRep336/00}
$S=2\int_{-\infty}^0\textrm{d}t\big[\big\langle
I(t)I(0)+I(0)I(t)-2\langle I\rangle^2\big\rangle\big]$. For
$|eV|>k_{\rm B}T$, the current noise is dominated by fluctuations
associated with the discrete nature of charge (shot noise), while
for low bias voltages, the thermal noise
dominates.~\cite{Blanter_PhysRep336/00} The general formula for
the current noise within the language of real-time diagrammatic
technique can be found in Ref.~[\onlinecite{Thielmann_PRL95/05}].

\section{Numerical results and discussion}

\begin{figure}[t]
  \includegraphics[width=0.7\columnwidth]{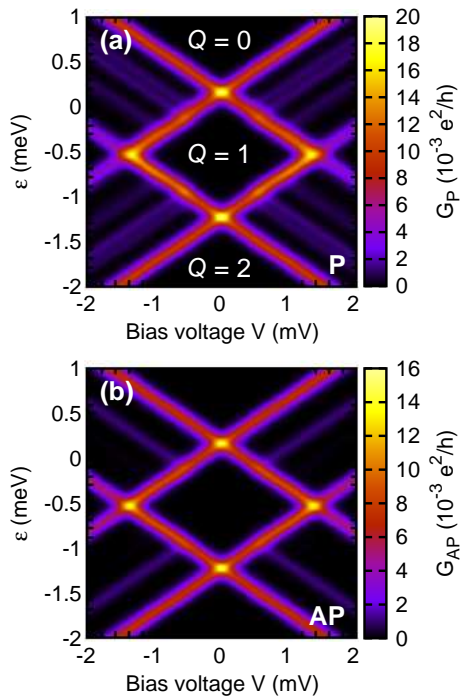}
  \caption{\label{Fig:2} (color online)
  The total (first plus second order)
  differential conductance in the parallel (a) and antiparallel
  (b) configurations for the parameters:
  $S=2$, $J=0.2$ meV, $D=0.05$ meV, $D_1=-0.005$ meV,
  $D_2=0.002$ meV, $U=1$ meV, $k_{\rm B}T=0.04$ meV, $P_L=P_R=0.5$,
  and $\Gamma=0.002$ meV.}
\end{figure}

In this section we present and discuss numerical results on charge
current, differential conductance, shot noise (Fano factor) and
tunnel magnetoresistance (TMR) in the linear and nonlinear
response regimes. The Fano factor
\begin{equation}
  F = \frac{S}{2e|I|} \,,
\end{equation}
describes deviation of the current noise from its Poissonian
value, $S_P = 2e|I|$, which is characteristic of uncorrelated in
time tunneling processes. On the other hand, the TMR is defined as
\cite{Julliere_PLA54/75, Barnas_PRL80/98,Weymann_PRB72/05_TMR}
\begin{equation}
  {\rm TMR} = \frac{I_{\rm P}-I_{\rm AP}}{I_{\rm AP}} \,,
\end{equation}
where $I_{\rm P}$ ($I_{\rm AP}$) is the current flowing through
the system in the parallel (antiparallel) magnetic configuration
at a constant bias voltage $V$. The TMR describes a change of
transport properties when magnetic configuration of the device
varies from antiparallel to parallel alignment -- the conductance
is usually larger in the parallel configuration and smaller in the
antiparallel one, although opposite situation is also possible.

Numerical results have been obtained for a hypothetical SMM
characterized by the spin number $S=2$ and strong uniaxial
magnetic anisotropy. However, we note that although in the
following we assume $S=2$, our considerations are still quite
general and qualitatively valid for molecules with larger spin
numbers. In fact, the choice of low molecule's spin allows us to
perform a detailed analysis of various molecular states mediating
the first and second-order tunneling processes. A large number of
molecular states for $S\gg 1$ would make the discussion rather
obscure. Apart from this, we assume a symmetrical coupling of the
molecule to the two external leads ($P_L=P_R=P$) and ferromagnetic
exchange coupling between the molecule's magnetic core and
electrons in the LUMO level. Later on, however, we will relax the
latter restriction and consider the situation where the exchange
coupling is antiferromagnetic. For clarity reasons, we disregard
the effects due to the negative sign of electron charge, i.e.
assume that charge current and particle (electron) current flow in
the same direction ($e>0$).

We start from some basic transport characteristics of the system
under consideration. In Fig.~\ref{Fig:2} we show the differential
conductance in the parallel and antiparallel configurations as a
function of the bias voltage and position of the LUMO level. The
latter can be experimentally changed by sweeping the gate voltage.
The density plot of the conductance displays the well-known
Coulomb diamond pattern. The average charge accumulated in the
LUMO level is $Q=\sum_\chi n(\chi)p_{\chi}^\textrm{st}$ (in the
units of $e$), where $n(\chi)=0,1,2$ denotes the number of
additional electrons on the molecule in the state $|\chi\rangle$.
When lowering energy of the LUMO level, the latter becomes
consecutively occupied with electrons. This leads to two peaks in
the linear conductance, separated approximately by $U$, which
correspond to single and double occupancy, respectively, see
Fig.~\ref{Fig:2} for $V=0$.

\begin{figure}[t]
  \includegraphics[width=0.7\columnwidth]{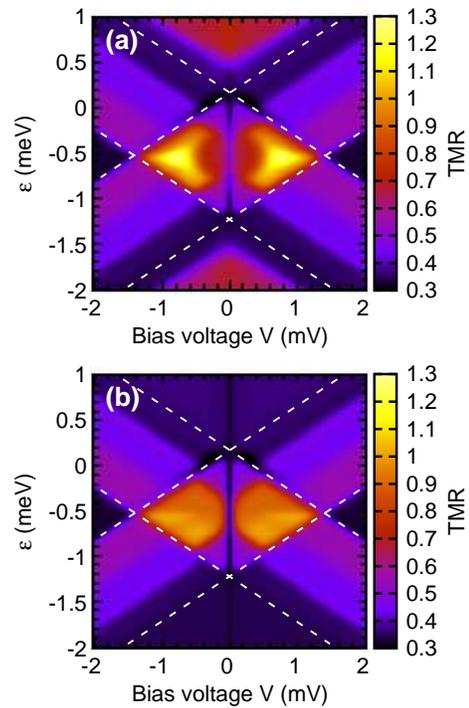}
  \caption{\label{Fig:3} (color online)
  Density plot of the total (first plus second order) TMR (a)
  and TMR calculated in the sequential tunneling approximation (b)
  plotted in the same scale and
  for the same parameters as in Fig.~\ref{Fig:2}.
  The sequential TMR is smaller than the total TMR.
  The dashed lines are only a guide for eyes,
  and they represent positions of the main conductance peaks,
  Fig.~\ref{Fig:2}, separating
  thus regions corresponding to different occupation states of the LUMO level.}
\end{figure}

Furthermore, in the nonlinear response regime, the differential
conductance shows additional lines due to tunneling through
excited states of the molecule. These features are visible in both
magnetic configurations. On the other hand, the hallmark of
spin-depended tunneling is the difference in magnitude of the
conductance in parallel and antiparallel configurations -- the
conductance in the parallel configuration is generally larger than
in the antiparallel one, see Fig.~\ref{Fig:2}. This difference is
due to spin asymmetry of tunneling processes, which leads to
suppression of the conductance when configuration changes from
parallel to antiparallel one.

The density plot of the TMR corresponding to Fig.~\ref{Fig:2} is
shown in Fig.~\ref{Fig:3}(a). As one can note, the magnitude of
TMR strongly depends on the transport regime. More precisely, TMR
can range from approximately ${\rm TMR} \approx P^2/(1-P^2) = 1/3$
(for $P=0.5$), which is characteristic of sequential tunneling
regime where all states of the LUMO level are active in transport,
\cite{Weymann_PRB72/05_TMR} to roughly twice the value resulting
from the Julliere model, \cite{Julliere_PLA54/75} ${\rm TMR}
\approx {\rm TMR^{Jull.}} = 4P^2/(1-P^2) = 4/3$, which can be
observed in the nonlinear response regime of the Coulomb blockade
diamond ($Q=1$), see Fig.~\ref{Fig:3}(a). For comparison, in
Fig.~\ref{Fig:3}(b) we display the TMR calculated using only the
sequential tunneling processes. One can see that the first-order
TMR is generally smaller than the total (first plus second order)
TMR. Furthermore, it is also clear that the second-order tunneling
processes modify TMR mainly in the Coulomb blockade regime ($Q=1$)
as well as in the cotunneling regimes where the LUMO level is
either empty ($Q=0$) or doubly ($Q=2$) occupied. On the other
hand, out of the cotunneling regime, the sequential processes
dominate transport and the role of second-order tunneling is
relatively small. As a consequence, the two results become then
comparable in these regions, see Fig.~\ref{Fig:3}(a) and
Fig.~\ref{Fig:3}(b).

\begin{figure}[t]
  \includegraphics[width=0.99\columnwidth]{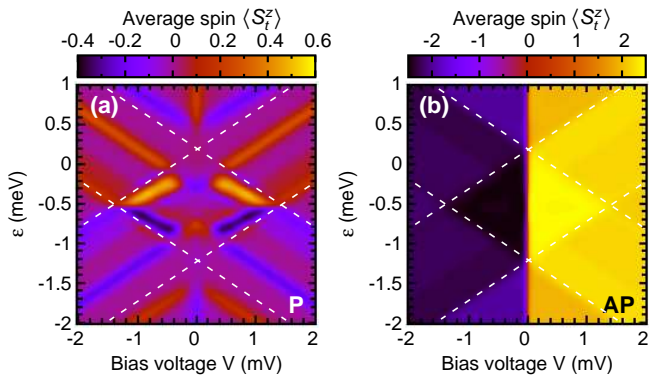}
  \caption{\label{Fig:4} (color online) The average
  value of the $z$th component of the molecules's total spin
  $\langle S_t^z\rangle$ for the parallel (a) and antiparallel
  (b) magnetic configurations.
  All parameters as in Fig.~\ref{Fig:2}.
  }
\end{figure}

Spin-dependent transport through a SMM has a significant impact on
its magnetic state. In Fig.~\ref{Fig:4} we show the average value
of the molecule's spin $z$th component in the stationary state,
$\langle S_t^z\rangle$, calculated as a function of the bias
voltage $V$ and energy of the  LUMO level $\varepsilon$. In the
antiparallel magnetic configuration, Fig.~\ref{Fig:4}(b), the
orientation of the molecule's spin is straightforwardly related to
the bias voltage, and for $V>0$ the spin is aligned along the easy
axis $+z$, whereas for $V<0$ it is aligned along the $-z$ axis.
Note, that in the regions corresponding to $Q=0$ and $Q=2$ the
spin is equal to that of magnetic core, while for $Q=1$ it also
includes the contribution from an electron in the LUMO level. By
contrast, in the parallel configuration, Fig.~\ref{Fig:4}(a), the
value of $\langle S_t^z\rangle$ in the stationary state can be
both positive and negative for each sign of the bias voltage, and
it varies in a rather limited range close to zero. Moreover,
$\langle S_t^z\rangle$ in the parallel (antiparallel) magnetic
configuration is an even (odd) function of the bias voltage $V$.

\begin{figure}[t]
  \includegraphics[width=0.99\columnwidth]{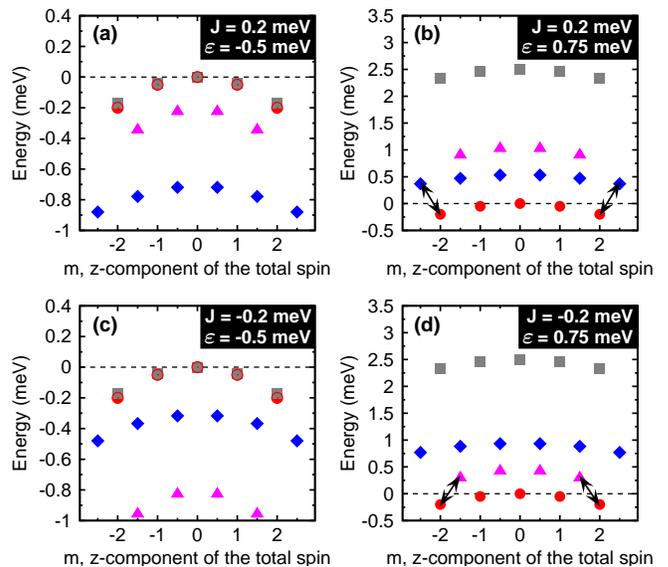}
  \caption{\label{Fig:5} (color online) Energy
  spectrum of the molecule under consideration
  (relevant parameters given in the caption of Fig.~\ref{Fig:2})
  for $\varepsilon=-0.5$ meV (a,c) and $\varepsilon=0.75$ meV (b,d)
  in the case of ferromagnetic (a)-(b) and
  antiferromagnetic (c)-(d) coupling between the SMM's core
  spin and the spin of electrons in the LUMO level.
  The dashed line represents the Fermi level of the leads
  when no external voltage bias is applied ($V=0$).
  Different sets of molecular states correspond to different
  values of the SMM's total spin $S_t$,
  and/or the occupation number of the LUMO level:
  $|2;0,m\rangle$ ({\large $\bullet$}), $|5/2;1,m\rangle$
  ({\scriptsize$\blacklozenge$}), $|3/2;1,m\rangle$ ($\blacktriangle$),
  and $|2;2,m\rangle$ ({\tiny $\blacksquare$}).
  Note that in (a) and (c) the degeneracy between states
  $|2;0,m\rangle$ and $|2;2,m\rangle$ takes place only for $m=0$.
  }
\end{figure}

To account for the transport properties in different regimes,
especially of TMR and shot noise, in the following we present and
discuss the gate and bias voltage dependence corresponding to
various cross-sections of the relevant density plots mentioned
above. More specifically, we will first consider transport
properties in the linear response regime
(Sec.~\ref{Sec:linear_response}), and then transport in the
nonlinear regime (Sec.~\ref{Sec:nonlinear_response}). In addition,
whenever advisable and possible, we will also compare and relate
our findings to existing results on quantum dot systems. At this
point, it is however worth emphasizing that the problem of
electron transport through a SMM is much more complex and
physically richer than in the case of single quantum dots.
\cite{Barnas_JPCM20/08} This is because now the transfer of
electrons occurs through many different many-body states of the
coupled LUMO level and molecule's magnetic core, see
Eq.~(\ref{eq:Hamiltonian_SMM}).

Since transport properties of a system are determined by its
energy spectrum, it is instructive to analyze it in more detail.
For molecules with only uniaxial anisotropy considered in this
paper, the molecule's Hamiltonian $\mathcal{H}_\textrm{SMM}$ can
be diagonalized analytically (the relevant formulas can be found
in Ref.~[\onlinecite{Misiorny_PSS_FA}], from where the notation
for molecular states has also been adopted). Energy spectrum of
the molecule under consideration is presented in Fig.~\ref{Fig:5}
for two different values of the LUMO level energy $\varepsilon$
and two values of the coupling parameter $J$. Each molecular state
$|S_t;n,m\rangle$ is labelled by the total spin number $S_t$, the
occupation number $n$ of the LUMO level, and the eigenvalue $m$ of
the $z$th component of the molecule's total spin, $S_t^z\equiv
S_z+\frac{1}{2}(c_\uparrow^\dagger c_\uparrow-c_\downarrow^\dagger
c_\downarrow)$, where the second term stands for the contribution
from electrons in the LUMO level. The change of the LUMO level
energy leads to the change in the energetic position of the
spin-multiplets $|5/2;1,m\rangle$, $|3/2;1,m\rangle$ and
$|2;2,m\rangle$ with respect to $|2;0,m\rangle$. The latter
multiplet corresponds to uncharged molecule and therefore is
independent of $\varepsilon$, see Fig.~\ref{Fig:5}.

\subsection{\label{Sec:linear_response} Transport in the linear response regime}

In this subsection we will focus on transport in the linear
response regime. As we have already mentioned above, conductance
in the linear response regime (see Fig. 2 for $V=0$), displays two
resonance peaks separated approximately by $U$. For $J>0$ and
$D(2S-1)\gg k_\textrm{B}T$, one can assume that the molecule is in
the spin states of lowest energy. The position of the conductance
peaks (resonances) corresponds then to $\varepsilon
=\varepsilon_{01}$,
    \begin{equation}\label{eq:eps_01}
    \varepsilon_{01} = \frac{JS}{2}+D_1S^2+\frac{g\mu_\textrm{B}|H_z|}{2},
    \end{equation}
for the transition from zero to single occupancy of the LUMO
level, and to $\varepsilon =\varepsilon_{12}$,
    \begin{equation}\label{eq:eps_12}
    \varepsilon_{12} = -\frac{JS}{2} -U +(D_1+D_2)S^2-\frac{g\mu_\textrm{B}|H_z|}{2},
    \end{equation}
for the transition from single to double occupancy. It is worth noting that the above expressions may be useful for estimating the
coupling constant $J$ from transport measurements. Moreover, from
the above formulas one can conclude that the middle of the Coulomb
blockade ($Q=1$ in Fig.~\ref{Fig:2}) regime corresponds to
$\varepsilon =\varepsilon_{m}$, with
\begin{equation}\label{eq:eps_m}
 \varepsilon_{m} = -\frac{U}{2} + \frac{2D_1+D_2}{2}S^2 \,,
\end{equation}
which for the parameters assumed in calculations gives
$\varepsilon_{m} = -0.516$ meV. Interestingly, $\varepsilon_{m}$
is independent of the exchange coupling $J$, anisotropy constant
$D$, and external magnetic field $H_z$, but it depends on the
Coulomb interaction $U$, corrections $D_1$ and $D_2$ to the
anisotropy due to finite occupation of the LUMO level, and the
molecule's spin number $S$. In fact, owing to finite constants
$D_1$ and $D_2$, the particle-hole symmetry is broken, which
manifests itself in an asymmetric behavior of transport
properties, as will be shown below.

Figure~\ref{Fig:6}(a) shows the total TMR in the linear response
regime, where for comparison TMR in the sequential transport
regime  is also displayed (dash-dotted line). Clearly, the results
obtained within the sequential tunneling approximation, which
yield a constant TMR equal to $P^2/(1-P^2)$, are not sufficient as
the total (first plus second order) linear TMR displays a
nontrivial dependence on the gate voltage. This behavior stems
from the dependence of the second-order processes on the
occupation number of the LUMO level.

\begin{figure}[t]
  \includegraphics[width=0.75\columnwidth]{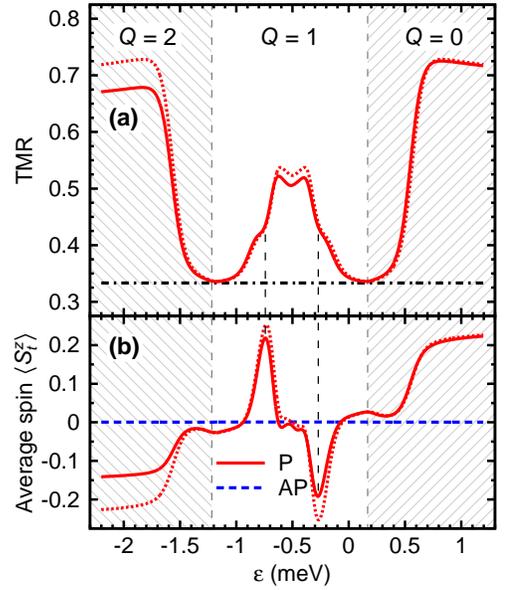}
  \caption{\label{Fig:6} (color online)
  (a) TMR in the linear response regime for
  the parameters as in Fig.~\ref{Fig:2} (solid line).
  The dot-dashed line shows the TMR calculated in
  the first-order approximation.
  (b) Average value of the $z$th component of the molecule's
  total spin in the (P) parallel (solid line) and
  (AP) antiparallel (dashed line) magnetic
  configurations. The dotted lines in (a) and (b) correspond to the case
  of $D_1=D_2=0$.}
\end{figure}

\begin{figure}[t]
  \includegraphics[width=0.85\columnwidth]{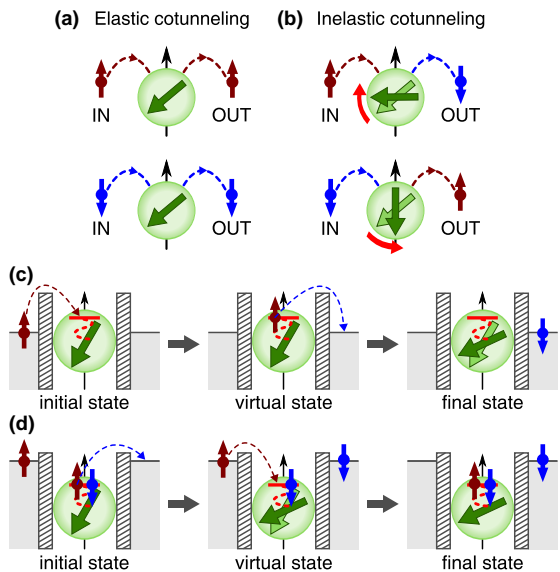}
  \caption{\label{Fig:7} (color online)
   (a)-(b) Schematic representation of the elastic and inelastic
   electron cotunneling processes. The two bottom panels
   show examples of inelastic cotunneling processes
   leading to increase of the $z$th component
   of the SMM's spin in the situation when
   the LUMO level is: (c) empty ($Q=0$) and (d) doubly occupied ($Q=2$).
   Note that $V\to 0$ in the linear response regime, so we assume $\mu_L =\mu_R +0^+$.
   When in the ground state the molecule is occupied by a single electron,
   $Q=1$, inelastic processes can generally occur via two virtual
   states associated with empty and doubly occupied LUMO level.
   }
\end{figure}

Generally, cotunneling processes can be divided into two groups
with respect to whether or not the molecule remains in its initial
state after a cotunneling process, Fig.~\ref{Fig:7}(a)-(b).
Although the cotunneling events do not change the charge state of
the molecule, they can, however, modify its spin state (inelastic
cotunneling). Moreover, the inelastic cotunneling processes can
lead to magnetic switching of the molecule's spin between two
lowest energy states, as shown schematically in
Fig.~\ref{Fig:7}(b). We note that in addition to {\it
double-barrier} cotunneling processes which transfer charge
between two different electrodes, there are also {\it
single-barrier} cotunneling processes, where an electron involved
in the cotunneling process returns back to the same electrode.
Although the latter processes do not contribute directly to the
current flowing through the system, they can affect all the
transport properties in an indirect way, by altering spin state of
the molecule.

\subsubsection{Cotunneling regime with empty and doubly occupied LUMO level}

When the LUMO level is either empty ($Q=0$) or fully occupied
($Q=2$), the total TMR in the corresponding cotunneling regions is
slightly larger than the Julliere value, \cite{Julliere_PLA54/75}
${\rm TMR^{Jull.}}=2P^2/(1-P^2)$ (${\rm TMR^{Jull.}}=2/3$ for
$P=0.5$), see Fig.~\ref{Fig:6}(a). Electron transport in these
regions is primarily due to elastic cotunneling processes which
change neither the electron spin in the LUMO level nor the spin of
molecule's core, and thus are fully coherent. An example of such
process is sketched in Fig.~\ref{Fig:7}(a). The enhancement of TMR
above the Julliere value is then associated with the exchange
coupling of the LUMO level to the molecule's core spin, which
additionally admits inelastic cotunneling processes in these
regions. In addition, the enhanced TMR may also result from the
fact that by using the {\it crossover} perturbation scheme,
\cite{Weymann_PRB72/05_TMR} we also include some effects
associated with third-order processes, which may further increase
the TMR. Moreover, unlike the case of a single quantum dot,
\cite{Barnas_JPCM20/08,Weymann_PRB72/05_TMR} the maximal values of
TMR reached for $Q=0$ and $Q=2$ do not necessarily have to be
equal, see Fig.~\ref{Fig:6}(a). Below, we discuss these new
features in more details.

From the energy spectrum displayed in Fig.~\ref{Fig:5}(b) follows
that the dominant elastic transfer of electrons between the leads
for $Q=0$ takes place via the following virtual transitions:
$|2;0,-2\rangle\leftrightarrow|5/2;1,-5/2\rangle$ and
$|2;0,2\rangle\leftrightarrow|5/2;1,5/2\rangle$ [indicated with
black arrows in Fig.~\ref{Fig:5}(b)]. In the parallel
configuration, the former transitions establish the transport
channel for minority electrons, whereas the latter ones for
majority electrons. The asymmetry between the occupation
probabilities of the states $|2;0,-2\rangle$ and $|2;0,2\rangle$
(with $|2;0,2\rangle$ being favored), which occurs due to
inelastic cotunneling processes, gives rise to increased transport
of majority electrons. On the other hand, there is no such
asymmetry in the antiparallel configuration. This, in turn, leads
to an enhancement of the TMR above the Julliere value,
Fig.~\ref{Fig:6}(a).

Similar arguments also hold for the case of $Q=2$, where the
molecular states $|2;2,m\rangle$ correspond to double occupancy of
the LUMO level. The main difference as compared to the situation
discussed above is that now in a cotunneling process the electron
first has to tunnel out of the LUMO level and then another
electron can tunnel onto the molecule [Fig.~\ref{Fig:7}(d)].
Analysis similar to that for $Q=0$ shows that in the parallel
configuration the inelastic cotunneling processes result in
lowering of the $z$th component of the SMM's spin, see
Fig.~\ref{Fig:6}(b). Moreover, the asymmetry between the
occupation probabilities of the states $|2;2,-2\rangle$ and
$|2;2,2\rangle$, where now $|2;2,-2\rangle$ is favored, leads to
increased elastic cotunneling of spin majority electrons and
therefore gives rise to enhanced TMR for $Q=2$.

Another interesting feature of TMR in the linear response regime,
shown in Fig. \ref{Fig:6}(a), is the difference in its magnitude
in the cotunneling regions corresponding to $Q=0$ and $Q=2$. This
is contrary to the case of Anderson model, where the linear TMR
was found to be symmetric with respect to the particle-hole
symmetry point, $\varepsilon = -U/2$. \cite{Weymann_PRB72/05_TMR}
In the case considered here, the situation is different due to
coupling of the LUMO level to the molecule's spin, and also due to
occupation dependent corrections to the anisotropy constant, see
Eq.~(\ref{eq:Hamiltonian_SMM}). These corrections reduce the
uniaxial anisotropy of the molecule with increasing number of
electrons in the LUMO level. As a result, the height of the energy
barrier between the two lowest molecular spin states is also
diminished for $Q=1$ and $Q=2$, and so are the energy gaps between
neighboring molecular states within the relevant spin multiplets.
For this reason, the probability distribution of the molecular
states for $Q=2$ (and also for $Q=1$) is more uniform than for
$Q=0$, see the solid line in Fig.~\ref{Fig:6}(b). Consequently,
the value of TMR for $Q=2$ is smaller than for $Q=0$. Thus, the
observed asymmetry with respect to $\varepsilon = \varepsilon_m$
is due to the lack of particle-hole symmetry in the system when
$D_1$ and $D_2$ are nonzero. However, if the influence of the LUMO
level's occupation on the anisotropy were negligible, $D_1\approx
D_2\approx0$ (the states $|2;0,m\rangle$ and $|2;2,m\rangle$ in
Fig.~\ref{Fig:5}(a) were then degenerate for every $m$), the
symmetry with respect to $\varepsilon =\varepsilon_m=-U/2$ would
be restored. This situation is presented by the dotted curves in
Fig.~\ref{Fig:6}, which clearly show that the asymmetric behavior
of TMR and $\langle S_t^z\rangle$ is related to the corrections to
anisotropy constants and the lack of particle-hole symmetry.

\subsubsection{Cotunneling regime with singly occupied LUMO level}

Interestingly, in the Coulomb blockade regime, with one electron
in the LUMO level ($Q=1$), the TMR reaches local maxima close to
the center of the Coulomb gap, and a shallow local minimum just in
the middle, i.e. for $\varepsilon=\varepsilon_m$. This behavior is
different from that observed in single-level quantum dots, where
linear TMR in the Coulomb blockade regime becomes suppressed and
reaches a global minimum when $\varepsilon=-U/2$.
\cite{Weymann_PRB72/05_TMR} As in the case of $Q=0$ and $Q=2$
discussed above, the origin of increased TMR for $Q=1$ can be
generally assigned to the modification of the probability
distribution of molecular states due to inelastic cotunneling
processes, Fig.~\ref{Fig:7}(b). In turn, the appearance of the
local minimum in the center of the $Q=1$ region is related to the
fact that when $\varepsilon = \varepsilon_m$, the virtual states
for leading inelastic cotunneling processes, which belong to spin
multiplets $|2;0,m\rangle$ and $|2;2,m\rangle$, become pairwise
degenerate (in the present situation, $|2;0,\pm2\rangle$ with
$|2;2,\pm2\rangle$). This means that in the parallel configuration
cotunneling processes involving empty and doubly occupied virtual
states occur at equal rates. As a consequence, the average spin on
the molecule tends to zero, see Fig.~\ref{Fig:6}(b), and TMR
displays a local minimum for $\varepsilon = \varepsilon_m$.

\subsubsection{Resonant tunneling regime}

For resonant energies, Eqs.~(\ref{eq:eps_01})-(\ref{eq:eps_12}),
where the occupancy $Q$ of the molecule changes, the sequential
tunneling processes play a dominant role. This results in the
reduction of TMR to approximately half of the Julliere value,
\cite{Julliere_PLA54/75} see the boundaries between the hatched
and non-hatched areas in Fig.~\ref{Fig:6}. The rate of first-order
tunneling processes increases whenever the two neighboring charge
states of the molecule become degenerate, provided that the
conditions $|\Delta n|=1$ and $|\Delta S_t^z|=1/2$ are
simultaneously satisfied, where $|\Delta n|$ and $|\Delta S_t^z|$
describe change in the occupation and spin of the molecule. This
means that for $\varepsilon =\varepsilon_{01}\approx 0.18$ meV the
degeneration between the empty and singly occupied states,
$|2;0;\pm2\rangle$ and $|5/2;1,\pm5/2\rangle$, is observed,
whereas for $\varepsilon =\varepsilon_{12}\approx -1.21$ meV the
states with a single and two electrons on the LUMO level,
$|5/2;1,\pm5/2\rangle$ and $|2;2;\pm2\rangle$, are degenerate.
Moreover, we also note that for $\Gamma \approx k_{B}T$, TMR can
be reduced further due to increased role of second-order processes
giving rise to the renormalization of the LUMO
level.~\cite{Weymann_PRB72/05_TMR}

\subsection{\label{Sec:nonlinear_response}Transport in the nonlinear response regime}

\begin{figure}[t]
  \includegraphics[width=0.99\columnwidth]{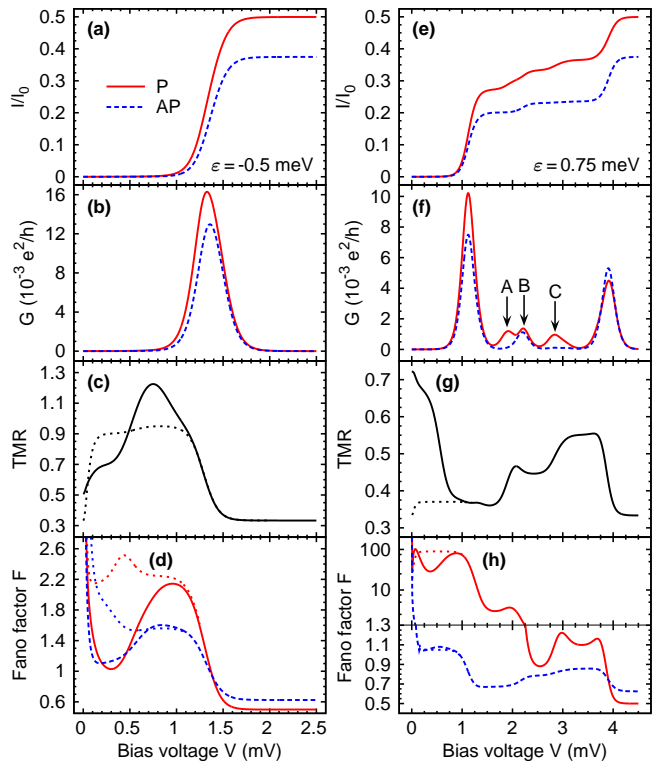}
  \caption{\label{Fig:8} (color online)
  Bias dependence of the current (a,e),
  differential conductance (b,f), Fano factor (d,h)
  in the parallel (solid lines) and antiparallel (dashed lines)
  configurations and TMR (c,g) for
  $\varepsilon = -0.5$ meV (a)-(d) and $\varepsilon=0.75$ meV (e)-(h).
  The parameters are the same as in Fig.~\ref{Fig:2}, and
  $I_0 = e\Gamma/\hbar \approx 0.5$ nA.
  The dotted lines show the results obtained
  taking into account only first-order tunneling processes.
  The effect of cotunneling is most pronounced
  in the TMR and Fano factor.}
\end{figure}

The influence of sequential tunneling on transport
characteristics, as well as on magnetic state of the SMM, grows
with increasing bias voltage. For voltages above the threshold for sequential tunneling,
first-order processes determine transport and the influence of
cotunneling is rather small. However, when applied voltage is
below the threshold, sequential tunneling becomes exponentially
suppressed and second-order processes give the dominant
contribution to the current, and need to be taken into account to
get a proper physical picture. Figure~\ref{Fig:8} shows the bias
dependence of the current, differential conductance, TMR and Fano
factor, calculated for $\varepsilon=-0.5$ meV and
$\varepsilon=0.75$ meV. The former case corresponds to the
situation where the LUMO level in equilibrium is singly occupied,
Fig.~\ref{Fig:5}(a), while in the latter case it is empty,
Fig.~\ref{Fig:5}(b). One can see that cotunneling significantly
modifies the first-order results in the blockade regimes and this
modification is most pronounced for TMR and shot noise.

\subsubsection{Transport characteristics in the case of $\varepsilon_{01}>\varepsilon>\varepsilon_{12}$}

Consider first the case when in equilibrium the LUMO level is
singly occupied (left panel of Fig.~\ref{Fig:8}). At low
temperatures and low voltages, the molecule with almost equal
probabilities is in one of the two ground states
$|5/2;1,\pm5/2\rangle$, Fig.~\ref{Fig:5}(a). When a small bias
voltage is applied, some current flows due to cotunneling
processes through virtual states of the system. If the bias
voltage exceeds threshold for sequential tunneling, the current
significantly increases and becomes dominated by first-order
processes, when electrons tunnel one-by-one through the molecule.

Since elastic cotunneling in the antiparallel configuration occurs
essentially through the minority-majority and majority-minority
channels, whereas for parallel alignment through the
majority-majority and minority-minority ones, one observes growth
of TMR with increasing bias voltage, which reaches a local maximum
just before the threshold for sequential tunneling. This is
associated with nonequilibrium spin accumulation in the LUMO level
for the antiparallel configuration, which leads to suppression of
charge transport and thus to enhanced TMR. Further increase of
transport voltage results in a decrease of TMR to approximately
1/3 (for $P_L=P_R=0.5$), which is typical of the sequential
tunneling regime, when all molecular states actively participate
in transport. \cite{Barnas_JPCM20/08,Weymann_PRB72/05_TMR} In the
parallel magnetic configuration all states are then equally
populated, so that average magnetic moment of the molecule
vanishes, $\langle S_t^z\rangle$=0. This differs from the
antiparallel case, in which only the states with large positive
$z$th component of the SMM's spin have remarkable probabilities.
Finally, we note that the slight shift between the peaks in
differential conductance corresponding to different magnetic
configurations, see Fig.~\ref{Fig:8}(b), is a consequence of
nonequilibrium spin accumulation in the LUMO level in the
antiparallel configuration. Similar behavior has been observed in
the case of transport through ferromagnetic single-electron
transistors. \cite{Weymann_PRB73/06}

The Fano factor in the parallel ($F_\textrm{P}$) and antiparallel
($F_\textrm{AP}$) configurations is shown in Fig.~\ref{Fig:8}(d).
For low bias voltages, the shot noise is determined by thermal
Johnson-Nyquist noise,  which results in a divergency of the Fano
factor for $V\to 0$ (current tends to zero). When a finite bias
voltage is applied to the system, the Fano factor in both magnetic
configurations drops to the value close to unity, which indicates
that transport occurs mainly due to elastic cotunneling processes.
Such processes are stochastic and uncorrelated in time, so the
shot noise is  Poissonian. When bias voltage increases further,
the shot noise is enhanced due to bunching of inelastic
cotunneling processes and reaches maximum just before threshold
for sequential tunneling. At the threshold voltage, sequential
tunneling processes begin to dominate transport and the noise
becomes sub-Poissonian. This indicates that tunneling processes in
the sequential tunneling regime are correlated due to Coulomb
correlation and Pauli principle, which generally gives rise to
suppressed shot noise as compared to the Poissonian value.
Furthermore, another feature clearly visible in the Coulomb
blockade regime is the difference in Fano factors for parallel and
antiparallel magnetic configurations. More specifically,  shot
noise in the parallel configuration is larger than in the
antiparallel one. This behavior is associated with the fact that
transport in the parallel configuration occurs mainly through two
competing majority-majority and minority-minority spin channels,
which in turn increases fluctuations, thus
$F_\textrm{P}>F_\textrm{AP}$.

\subsubsection{Transport characteristics in the case of $\varepsilon>\varepsilon_{01}$}

Let us consider now the situation shown in the right panel of
Fig.~\ref{Fig:8}, i.e. when the  LUMO level of the molecule is
empty  at equilibrium, Fig.~\ref{Fig:5}(b). The initial large
value of TMR, whose origin was discussed above, drops sharply as
the bias voltage approaches the threshold value for sequential
transport. In turn, the first pronounced peak in differential
conductance appears when the following transitions become allowed:
$|2;0,\pm2\rangle\leftrightarrow|5/2;1,\pm5/2\rangle$ [denoted by
arrows is Fig.~\ref{Fig:5}(b)]. It is important to note that, when
a spin-multiplet enters the transport energy window, the first
states that take part in transport are those with the largest
$|\langle S_t^z\rangle|$ (lowest energy). Consequently, in the
parallel magnetic configuration the system can be temporarily
trapped in some molecular spin states of lower energy. For larger
bias voltage, additional small peaks appear in the conductance for
parallel configuration, and some of them are also visible in the
antiparallel configuration. In general, these peaks are related to
transitions involving states from the multiplet $|3/2;1,m\rangle$:
$|2;0,\pm1\rangle\leftrightarrow|3/2;1,\pm3/2\rangle$ (A),
$|2;0,\pm2\rangle\leftrightarrow|3/2;1,\pm3/2\rangle$ (B) and
$|3/2;1,\pm3/2\rangle\leftrightarrow|2;2,\pm2\rangle$ (C),
respectively, see Fig.~\ref{Fig:8}(f). In the parallel
configuration all the three peaks are visible, whereas for
antiparallel alignment only the peak B can be clearly resolved.
Since in the antiparallel configuration tunneling processes tend
to increase the $z$th component of the SMM's total spin, the
probability of finding the molecule in any of the spin states
$|2;0,m\rangle$ differs significantly from zero only for $m=2$. As
a consequence, in the antiparallel configuration most favorable
transitions are those having the initial state $|2;0,2\rangle$,
and thus the peaks A and C are suppressed, see
Fig.~\ref{Fig:8}(f).

Furthermore, as soon as all states within a certain spin-multiplet
become energetically accessible, the probability of finding the
molecule in each of these states becomes roughly equal. On the
other hand, in the antiparallel configuration the system tends
towards maximum value (for $V>0$) of the $z$th component of SMM's
spin. For these reasons,  some regions of the increased TMR are
present in Fig.~\ref{Fig:8}(g).

The corresponding Fano factor is shown in Fig.~\ref{Fig:8}(h). At
low bias, the Fano factor drops with increasing voltage. However,
its bias dependence is distinctively different in both magnetic
configurations. In the antiparallel configuration, the Fano factor
tends to unity, indicating that transport is due to uncorrelated
tunneling events. In the parallel configuration, on the other
hand, we observe large super-Poissonian shot noise. The increased
current fluctuations result mainly from the interplay between
different cotunneling processes and bunching of inelastic
cotunneling. In addition, as mentioned previously, in the parallel
configuration the molecule can be temporarily trapped in some
molecular spin states of lower energy, which also gives rise to
super-Poissonian shot noise. When the bias voltage is increased
above the threshold for sequential tunneling, the Fano factor
becomes suppressed and the shot noise is generally sub-Poissonian.
Finally, we also note that super-Poissonian shot noise in the
cotunneling regime has already been observed in quantum dots and
carbon nanotubes,
\cite{Cottet_PRL2004,Onac_PRL2006,Zhang_PRL2007,Barnas_JPCM20/08}
where the increased noise was associated with bunching of
inelastic spin-flip cotunneling events.

\subsection{\label{Sec:different_J}Dependence on exchange coupling strength }

\begin{figure}[t]
  \includegraphics[width=0.99\columnwidth]{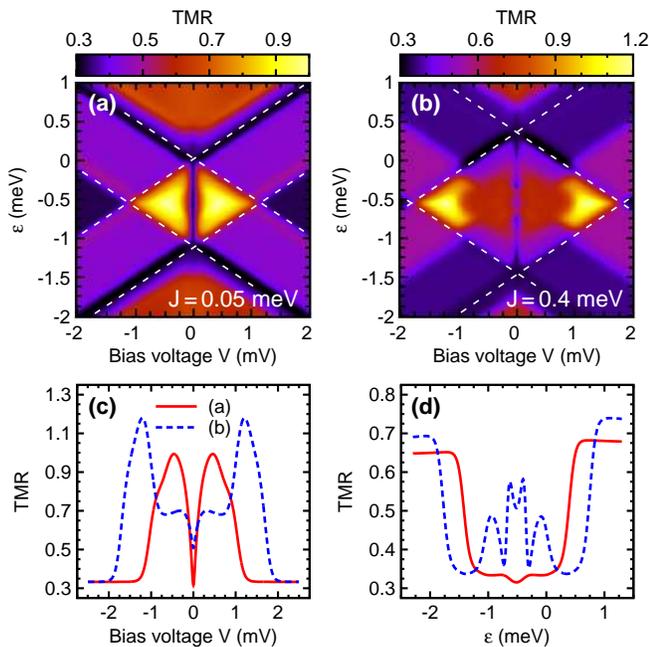}
  \caption{\label{Fig:9} (color online)
  Density plot of TMR for $J=0.05$ meV (a) and $J=0.4$ meV (b).
  The bottom panels display cross-sections
  of (a) and (b) for the constant energy of
  the molecule's LUMO level $\varepsilon=-0.5$ meV
  (c) and the constant bias voltage $V=0$ mV (d).
  Other parameters are the same as in Fig.~\ref{Fig:2}.}
\end{figure}

Tunnel magnetoresistance may become significantly changed by
altering the strength of ferromagnetic exchange coupling between
the LUMO level and the SMM's core spin, as shown in
Fig.~\ref{Fig:9}. With decreasing $J$, the energy separation
between the relevant molecule states corresponding to the single
occupancy of the LUMO level is also diminished (slanted squares
and triangles in Fig.~\ref{Fig:5} start then approaching each
other). This, in turn, leads to a reduction in size of the central
diamond-shaped region, representing transport in the Coulomb
blockade regime through the molecule with one electron in the LUMO
level. As follows from Fig.~\ref{Fig:9}(a), behavior of TMR for
small values of $J$ starts bearing some resemblance to that of a
single-level quantum
dot.~\cite{Barnas_JPCM20/08,Weymann_PRB72/05_TMR} Furthermore, in
the linear response regime, Fig.~\ref{Fig:9}(d), the enhanced TMR
around the electron-hole symmetry point is no longer visible, and
instead a global minimum develops there. In fact, in the limit of
$J=0$ one observes a simple quantum-dot-like transport behavior.
\cite{Weymann_PRB72/05_TMR,Barnas_JPCM20/08}

In the opposite limit of large $J$ shown in Fig.~\ref{Fig:9}(b),
the maxima in the total linear TMR are shifted away from the zero
bias point. This is a consequence of increased energy gaps between
the ground states $|5/2;1,\pm5/2\rangle$ and the nearest lying
states satisfying $|\Delta n|=1$ and $|\Delta S_t^z|=1/2$, i.e.
$|2;0(2),\pm2\rangle$. Another interesting feature of TMR visible
in the linear response regime is the presence of additional two
local minima around $\varepsilon=\varepsilon_m$, see
Fig.~\ref{Fig:9}(d). Some precursors of these minima can be
actually seen also in Fig.~\ref{Fig:6}(a) as two steep steps on
both sides of the plot's central part. Generally, they stem from
an uneven probability distribution of the molecular spin states
with positive and negative $z$th component of the SMM's spin in
the parallel magnetic configuration, see Fig.~\ref{Fig:6}(b). This
in turn means that elastic cotunneling processes occur mainly
through the minority-minority spin channel, so that transport is
effectively suppressed. In the present situation, the minima are
more distinct due to larger energy separation between the
spin-multiplets $|5/2;1,m\rangle$ and $|3/2;1,m\rangle$.

In the nonlinear response regime, on the other hand, the TMR
exhibits a minimum at zero bias and starts increasing with the
bias voltage to reach a maximum around the threshold for
sequential tunneling. This is associated with nonequilibrium spin
accumulation in the LUMO level, which is present in the
antiparallel configuration. We note that although the magnitude
and position of the TMR maxima in the nonlinear response regime of
the Coulomb blockade depend significantly on the exchange constant
$J$, the general qualitative behavior of TMR is rather independent
of $J$, see Figs.~\ref{Fig:9}(c) and \ref{Fig:8}(c).

\subsection{\label{Sec:magnetic_field}Transport in the presence of a longitudinal external magnetic field}

\begin{figure}[t]
  \includegraphics[width=0.99\columnwidth]{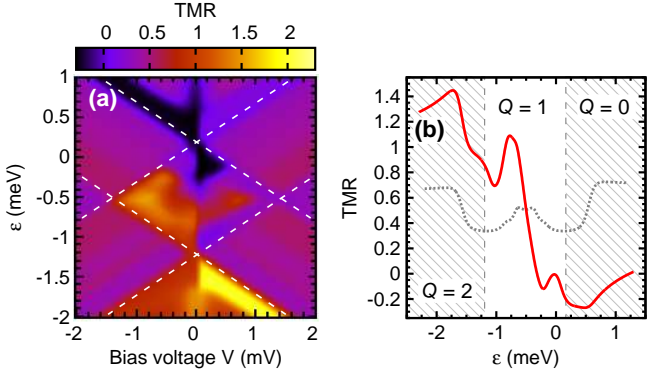}
  \caption{\label{Fig:10} (color online)
  (a) Density plot of TMR
  in the case when the external magnetic field $H_z=0.216$ T
  ($g\mu_\textrm{B}H_z=0.025$ meV) is applied along
  the easy axis of the molecule.
  (b) TMR in the linear response regime (solid line).
  For comparison, TMR in the absence of external
  magnetic field (dotted line in (b)) is also shown.
  The other parameters are the same as in Fig.~\ref{Fig:2}.}
\end{figure}

Let us consider now the main effects due to a finite magnetic
field applied to the system. When the field is along the easy axis
of the molecule, its effects occur \emph{via} modification of the
energy of molecular spin states. On the other hand, when the field
possesses also a transversal component, it leads to
symmetry-breaking effects and the $z$th component of the SMM's
total spin is no more a good quantum number.~\cite{Timm_PRB76/07}
If the magnetic field is additionally time-dependent, one can
expect the phenomenon of quantum tunneling of magnetization to
occur. \cite{Chudnovsky_book_MQT, Gatteschi_AngewChemIntEd42/03,
Misiorny_PSS_FA} Since the primary focus of the present paper is
on transport through SMMs with uniaxial anisotropy, in the
following we consider only a longitudinal magnetic field.

\begin{figure}[t]
  \includegraphics[width=0.99\columnwidth]{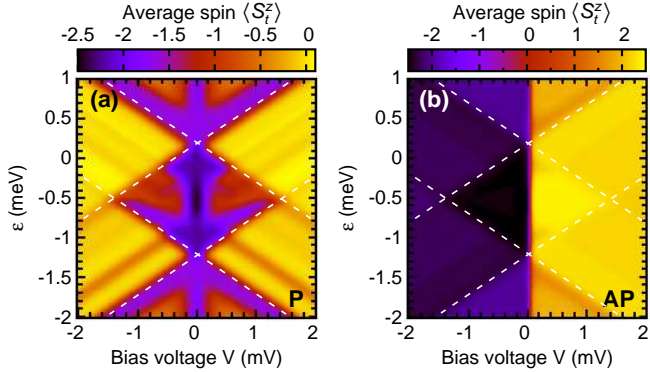}
  \caption{\label{Fig:11} (color online)
  Average value of the $z$th component of the
  total molecule's spin in the parallel (a) and antiparallel
  (b) magnetic configurations,
  when an external field $H_z=0.216$ T is applied
  along the $z$-axis.
  The other parameters are the same as in Fig.~\ref{Fig:2}.}
\end{figure}

Figure~\ref{Fig:10}(a) shows the density plot of TMR for a
magnetic field applied along the easy axis of a SMM. Despite
rather modest value of the field (for comparison, in the
experiment on the $\textrm{Mn}_{12}$ molecule attached to
nonmagnetic metallic electrodes by Jo {\it et al.}, the field of 8
T was used, Ref.~[\onlinecite{Jo_NanoLett6/06}]), a drastic change
in transport properties of the system is observed [contrast
Fig.~\ref{Fig:10}(a) with Fig.~\ref{Fig:3}(a)]. First, the field
breaks the symmetry with respect to the bias reversal. Second, it
admits the situation when conductance in the antiparallel magnetic
configuration is larger than in the parallel one (black regions
corresponding to negative TMR). Furthermore, in the parallel
configuration the average spin $\langle S_t^z\rangle$ in the
Coulomb blockade region can take large negative values, while in
the absence of magnetic field the SMM's spin prefers orientation
in the plane normal to the easy axis. This implies that for
parallel alignment of leads' magnetizations, the molecule's spin
tends to align antiparallel to the $z$-axis, Fig.~\ref{Fig:11}(a).
However, when the sequential tunneling processes are allowed, this
tendency is generally reduced. In the antiparallel configuration,
on the other hand, the behavior of the average molecule's spin is
similar to that for $H_z=0$, see Figs.~\ref{Fig:11}(b) and
\ref{Fig:4}(b).

\begin{figure}[t]
  \includegraphics[width=1\columnwidth]{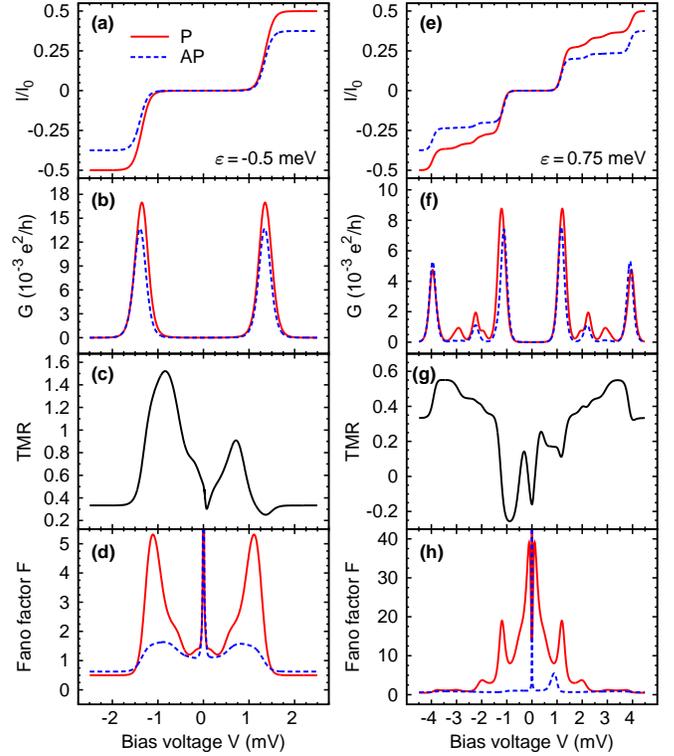}
  \caption{\label{Fig:12} (online color)
  The current (a,e),
  differential conductance (b,f), Fano factor (d,h)
  in the parallel (solid lines) and antiparallel (dashed lines)
  configurations, and the TMR (c,g) for
  $\varepsilon = -0.5$ meV (a)-(d) and $\varepsilon=0.75$ meV (e)-(h)
  as a function of the bias voltage.
  An external magnetic field $H_z=0.216$ T
  is applied along the $z$-axis, while
  the other parameters are the same as in Fig.~\ref{Fig:2}.
}
\end{figure}

In the linear response regime, a large change of TMR is observed
when $\varepsilon$ is comparable to $\varepsilon_m$, i.e. in the
middle of the Coulomb blockade regime, see Fig.~\ref{Fig:10}(b).
This stems from the fact that at this point the dominating
spin-dependent channel for transport due to cotunneling processes
in the parallel magnetic configuration switches from the
minority-minority channel (for $\varepsilon>\varepsilon_m$) to
majority-majority one (for $\varepsilon<\varepsilon_m$). In the
antiparallel configuration, on the other hand, the dominant
channel is rather associated with majority-minority spin bands,
irrespective of the position of the LUMO level. As a consequence,
for $\varepsilon>\varepsilon_m$ the current in the parallel
configuration is smaller than that in the antiparallel one,
leading to negative TMR, whereas for $\varepsilon<\varepsilon_m$
the situation is opposite and one finds a large positive TMR
effect, see Fig.~\ref{Fig:10}(b).

The transport characteristics in the nonlinear response regime,
and  in the presence of external magnetic field, are shown in
Fig.~\ref{Fig:12}, where the left (right) panel corresponds to the
situation where in the ground state the molecule is singly
occupied (empty). The asymmetry with respect to the bias reversal
is clearly visible, especially in the tunnel magnetoresistance,
see Fig.~\ref{Fig:12}(c) and (g). Interestingly, this asymmetric
behavior is mainly observed in the cotunneling regime, as can be
also seen in Fig.~\ref{Fig:10}(a). This results from the fact that
the degeneracy of the molecule's ground state is removed for $H_z
\neq 0$ and the SMM becomes polarized. In turn,  transport in the
cotunneling regime depends mainly on the system's ground state,
which is the initial state for the cotunneling processes.  As a
consequence, in the parallel configuration the current is always
mediated by electrons belonging to the same spin bands of the
leads, whereas in the antiparallel configuration, the dominant
transport channel is associated either with majority or minority
electrons, depending on the direction of the current flow. Thus,
the current in the antiparallel configuration becomes in general
asymmetric with respect to the bias reversal, which gives rise to
the  associated asymmetric behavior of TMR.

For voltages larger than the splitting due to the Zeeman term
($g\mu_\textrm{B}H_z=0.025$ meV), the inelastic cotunneling
processes start taking part in transport. The competition between
the elastic and inelastic cotunneling leads in turn to large
super-Poissonian shot noise, which in the parallel configuration
is enhanced due to additional fluctuations associated with
cotunneling through majority-majority and minority-minority spin
channels, see Fig.~\ref{Fig:12}(d) and (h). On the other hand,
when the voltage exceeds threshold for sequential tunneling, more
states take part in transport and the asymmetry with respect to
the bias reversal is suppressed. The same tendency is observed in
the shot noise, which in the sequential tunneling regime becomes
generally sub-Poissonian.

\subsection{\label{Sec:anitferromagnetic_coupling}Antiferromagnetic coupling between
the LUMO level and  SMM's core spin}

\begin{figure}[t]
  \includegraphics[width=0.99\columnwidth]{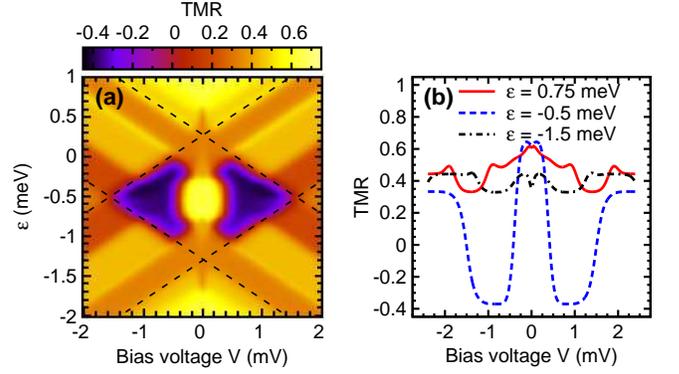}
  \caption{\label{Fig:13} (color online)
  (a) The total tunnel magnetoresistance
  in the case of antiferromagnetic coupling between the SMM's
  core spin and the spin in the LUMO level, calculated for
   $J=-0.2$ meV and other parameters as in Fig.~\ref{Fig:2}.
  (b) Representative cross-sections of the density
  plot in (a) for several values of the LUMO level energy $\varepsilon$.}
\end{figure}

The numerical results presented up to now concerned the case of
ferromagnetic coupling ($J>0$) between the LUMO level and the
SMM's core spin. However, since the type of such an interaction
generally depends on the SMM's internal structure, the exchange
coupling  can be also of antiferromagnetic type ($J<0$). In this
subsection we discuss how the main transport properties of the
system change when the exchange coupling becomes
antiferromagnetic.

First of all, we note that in the case of antiferromagnetic
coupling between the LUMO level and molecule's core spin the
formulas estimating the position of conductance resonances need
some modification. Equations~(\ref{eq:eps_01})-(\ref{eq:eps_12})
were derived assuming the degeneracy between the states
$|2;0,\pm2\rangle$ ($|5/2;1,\pm5/2\rangle$) and
$|5/2;1,\pm5/2\rangle$ ($|2;2,\pm2\rangle$). For $J<0$, however,
the condition has to be modified by changing
$|5/2;1,\pm5/2\rangle$ into $|3/2;1,\pm3/2\rangle$, where the
upper signs apply for $H_z<0$, and the lower ones for $H_z>0$. The
relevant equations take now the following form:
    \begin{equation}
    \varepsilon_{01}=\frac{|J|}{4}+D_1S^2 - \Delta\varepsilon
    \end{equation}
for the transition from  empty to singly occupied states, and
    \begin{equation}
    \varepsilon_{12}=-\frac{|J|}{4} - U +(D_1+D_2)S^2 +
    \Delta\varepsilon
    \end{equation}
for the transition between singly and doubly occupied states,
where
    \begin{align}
    \hspace*{-0.1cm}
    \Delta\varepsilon&=D^{(1)}\frac{2S-1}{2} + \frac{g\mu_\textrm{B}|H_z|}{2}
    \nonumber\\
    &-\sqrt{D^{(1)}(D^{(1)}+|J|)\frac{(2S-1)^2}{4}+\frac{J^2}{16}(2S+1)^2},
    \end{align}
with $D^{(1)}=D+D_1$.

The most apparent new feature of the total TMR for $J<0$, as shown
in Fig.~\ref{Fig:13}(a), is its negative value in the Coulomb
blockade regime ($Q=1$). The negative TMR occurs in transport
regimes where the maximum of TMR was observed for $J>0$, i.e.
close to the threshold for sequential tunneling, see
Fig.~\ref{Fig:3}(a). Such behavior of TMR originates from the fact
that now spin-multiplets $|5/2;1,m\rangle$ and $|3/2;1,m\rangle$
exchange their positions, Fig.~\ref{Fig:5}(c)-(d), so that the
multiplet corresponding to smaller total spin of the molecule for
antiferromagnetic coupling corresponds to lower energy.
Consequently, in the Coulomb blockade the current flowing in the
antiparallel configuration is larger than that in the parallel
configuration, which gives rise to the negative TMR effect.

\begin{figure}[t]
  \includegraphics[width=0.75\columnwidth]{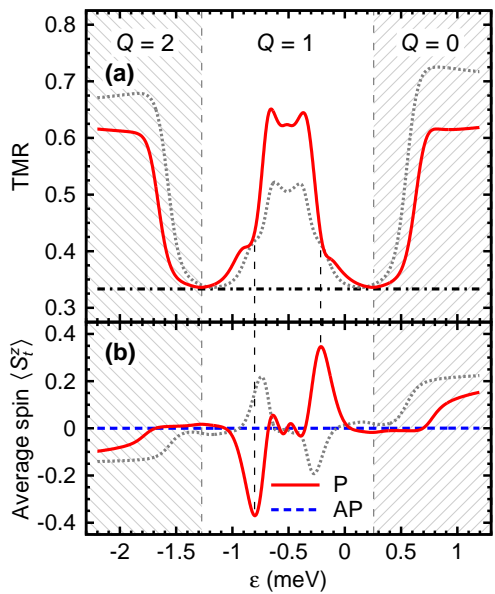}
  \caption{\label{Fig:14} (color online)
  Tunnel magnetoresistance (a) and the $z$th
  component of the molecule's total spin (b)
  calculated in the linear response regime
  for the antiferromagnetic coupling of the SMM's
  core spin with the spin of the LUMO level
  ($J=-0.2$ meV and other parameters as in Fig.~\ref{Fig:2}).
  Dotted lines show the results obtained for
  the case of ferromagnetic exchange coupling,
  see Fig.~\ref{Fig:6} -- in (b) the dotted line
  corresponds to the parallel configuration.}
\end{figure}

The linear response TMR is shown in Fig.~\ref{Fig:14}(a). Unlike
the case of ferromagnetic coupling, the values of TMR for $Q=0$
and $Q=2$ are smaller as compared to those in the case of
transport through single-level quantum dots.
\cite{Barnas_JPCM20/08, Weymann_PRB72/05_TMR} On the other hand,
for $Q=1$ the TMR can take values exceeding those found in the
case of ferromagnetic exchange coupling. For
$\varepsilon>\varepsilon_{01}$, the equilibrium probability
distribution of different molecular spin states $|2;0,m\rangle$
becomes changed owing to inelastic cotunneling processes,
similarly as described in Sec.~\ref{Sec:linear_response}. The key
difference with respect to $J>0$ is that now dominating elastic
cotunneling transitions  for $Q=0$ are those with initial states
$|2;0,\pm2\rangle$ and virtual states $|3/2;1,\pm3/2\rangle$
[indicated by black arrows in Fig.~\ref{Fig:5}(d)].

\section{Summary and conclusions}

We have systematically analyzed the transport properties of a
single-molecule magnet coupled to ferromagnetic leads in both
sequential and cotunneling regimes. The transport processes in
such a system occur due to tunneling through the LUMO level which
is exchange coupled to the molecule's core spin. By employing the
real-time diagrammatic technique, we have calculated the current,
tunnel magnetoresistance and shot noise in both the linear and
nonlinear response regimes. The results show that the inclusion of
second-order processes is crucial for a proper description of
transport characteristics.

Assuming the ferromagnetic coupling between the LUMO level and the
molecule's core spin, we have shown that TMR in the Coulomb
blockade regime can be enhanced above the Julliere value. This
enhancement is associated with nonequilibrium spin accumulation in
the molecule. Moreover, we have found an asymmetric behavior of
the linear response TMR with respect to the middle of the Coulomb
blockade regime, and its strong dependence on the number of
electrons in the LUMO level. The asymmetry is associated with
corrections to anisotropy constant due to a nonzero occupation of
the molecule, which breaks the particle-hole symmetry in the
system. In addition, we have shown that the competition between
the elastic and inelastic second-order processes leads to large
super-Poissonian shot noise. The shot noise is further enhanced in
the parallel configuration due to additional fluctuations
associated with majority-majority and minority-minority spin
channels for electronic transport. On the other hand, for bias
voltages above the threshold for sequential tunneling, the shot
noise becomes generally sub-Poissonian, indicating the role of
correlations in sequential transport.

We have also discussed how transport properties depend on the
strength of the exchange coupling $J$ between the LUMO level and
the molecule's core spin. When the exchange coupling is relatively
weak, the transport behavior of the system resembles that of
single-level quantum dots, whereas with increasing exchange
constant, the transport characteristics change in a nontrivial way
and become distinctively different from those of quantum dots. In
addition, it turned out that the position of maxima of TMR in the
Coulomb blockade depend linearly on the strength of the exchange
coupling. This may be useful in determining the magnitude of
exchange constant experimentally.

Furthermore, we have studied the effects of external magnetic
field and shown that current flowing through the SMM becomes then
asymmetric with respect to the bias reversal. We have found a
strong dependence of TMR on the number of electrons occupying the
LUMO level. When the LUMO level is empty, the TMR may become
negative, while for doubly occupied LUMO level tunnel
magnetoresistance is much enhanced. Finally, we have also
discussed how transport properties change when the coupling
between the LUMO level and molecule's core becomes
antiferromagnetic. In that case we predict a large negative TMR
effect in the Coulomb blockade regime, exactly where for
ferromagnetic coupling an enhanced TMR was observed. Thus, the
sign of TMR may provide an information on the type of  exchange
interaction, which may be of assistance for future experiments.

To conclude, we note that although the numerical results presented
in this paper concern SMMs coupled to ferromagnetic leads, most of
the qualitative results are applicable also to SMMs coupled to
nonmagnetic electrodes. Apart from this, we note that the model we
have studied also corresponds to systems consisting of a
single-level quantum dot exchange-coupled to a spin $S$. In fact,
very recently a similar device built of a quantum dot coupled
through spin exchange interaction to metallic island have been
implemented to experimentally access the quantum critical point
between the Fermi liquid and non-Fermi liquid regimes.
\cite{Potok_Nature2007}


\begin{acknowledgments}

This work, as part of the European Science Foundation EUROCORES
Programme SPINTRA, was supported by funds from the Ministry of
Science and Higher Education as a research project in years
2006-2009 and the EC Sixth Framework Programme, under Contract N.
ERAS-CT-2003-980409. M.M. was also supported by the Adam
Mickiewicz University Foundation and by funds from the Ministry of
Science and Higher Education as a research project in years
2008-2009. I.W. acknowledges support from the Ministry of Science
and Higher Education through a research project in years 2008-2010
and the Foundation for Polish Science.

\end{acknowledgments}



\end{document}